# Methods of Repairing Virus Infected Files
## A TRIZ based analysis

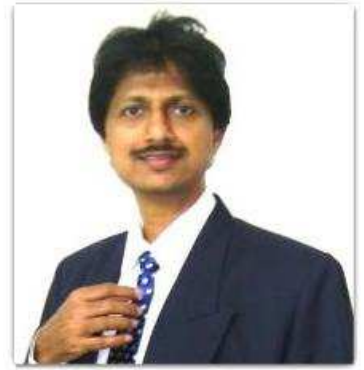

**By- Umakant Mishra, Bangalore, India**

umakant@trizsite.tk, http://umakant.trizsite.tk

**Contents**

# 1. What is virus infection

A computer virus is a manmade destructive computer program or code. One of the fundamental characteristics of a virus is that it replicates its code to other programs or computers. This process of replication is called infection. In order to replicate its code, generally it has to explore a suitable host program (such as, .exe for macro viruses or .doc for macro viruses) and then copy its code to the host program. The infected program again finds another victim to infect further.

Before discussing about virus cleaning, it will be important to discuss how a virus infects a program. A typical virus generally (although not in all the cases) adds itself to the end of a program. In that case the size of the host program increases because of this addition of this extra (viral) code. In order to execute this viral code, the virus overwrites the first bytes of the file with a "jump" instruction which makes the execution jump to the viral code. After the viral code is executed, the virus repairs the first few bytes overwritten by the virus in order to return control to the original file. Thus a typical virus (not in case of a worm or Trojan) has to depend on a host program to survive and operate. Some interesting facts about virus infection are as below.

> During the process of infection the content of the host file (or the file that is infected) has to change in order to include the virus code. **Every file must change during infection. In other words, there is no possibility of a file being infected but not changed.**
>
> In most cases the infected host program still works after infection. In those cases the size of the host program must have increased (except in case of cavity viruses), because in this case the size of the infected file will be equal to the size of the original file + size of the virus code.
>
> If the size of the host file is not increased after infection, then the virus must have overwritten some part of the file (except in case of cavity viruses). These cases are either irreparable or very difficult to repair.

### 1.1 Types of virus infections

As there are different types of viruses, their methods of infections are also different. Some viruses infect the boot sector and partition table (boot sector viruses). Some viruses remain in memory all the time (memory resident viruses) since the computer is switched on. Some viruses remain in the body of executable files (file viruses), or document files (macro viruses) or other type of files. Some viruses infect emails (email viruses) and travel through computer networks. Some viruses are complete programs and work independently without depending on other files (worms, trojans). Some viruses take over actions of operating systems (rootkits) thereby causing the system to malfunction.

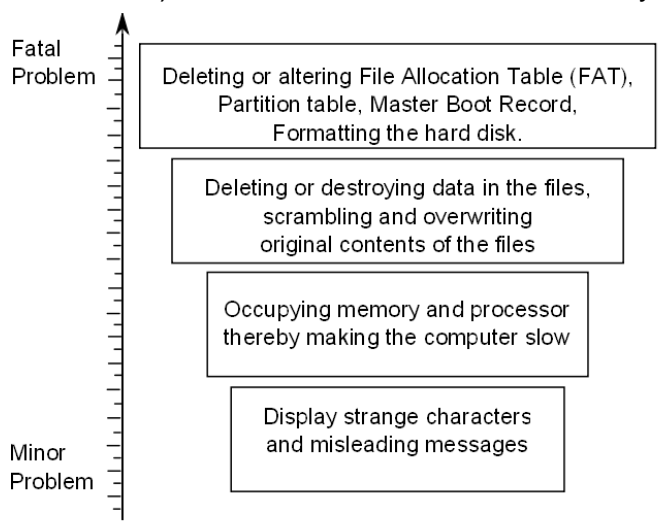

Different levels of harms made by computer viruses

Every infection does not cause same level of damage. Some infections cause very mirror disturbances while other infections cause moderate to high level of damages. The dangerous viruses may cause serious damages like formatting the hard disk or destroying the data etc. thereby making the computer unusable.

**1.2 Patterns of virus infections**

There are several patterns of virus infections. Most viruses append themselves to the host program and modify the header of the original host so that the execution will begin at the virus code rather than the original host code. Some viruses add themselves at the beginning of the host program. These are the two simple patterns where the original program code remains in one single block even after the infection.

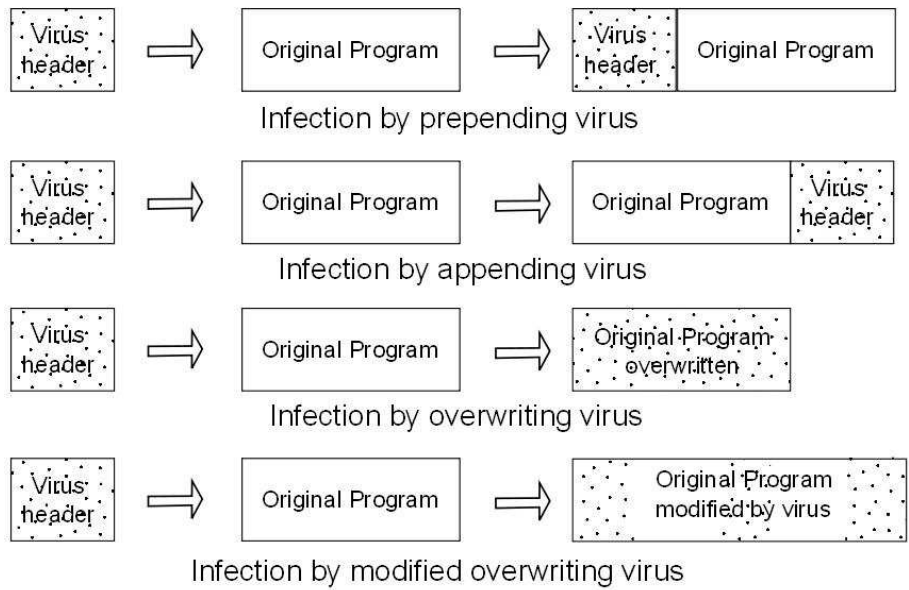

However there are more complicated patterns too. The overwriting viruses overwrite a portion of the host program and modify the header of the host program so as to begin the execution inside the virus. In this case the size of the infected file may not change. This pattern is dangerous as the original program suffers from permanent damage. Modified overwriting viruses write themselves at the beginning, end and other places in the original program. There are many complicated modified overwriting viruses who scramble and overwrite the original host in various different ways and make them useless after infection. (Note: all these infections patterns are relevant for typical file viruses. The infection patterns of worms, Trojans, rootkits or even macro viruses are different.)

## 2. What actions are taken on an infected file

Although an anti-virus aims at preventing any virus attack, there are chances that some files are infected earlier to anti-virus installation or during a period when the virus signatures are not updated. In that case, the anti-virus must disinfect the infected files. This is one of the difficult and most important functions of any anti-virus program. The anti-virus has to apply various methods in order to remove the virus code from the infected file and restore the file in its original form.

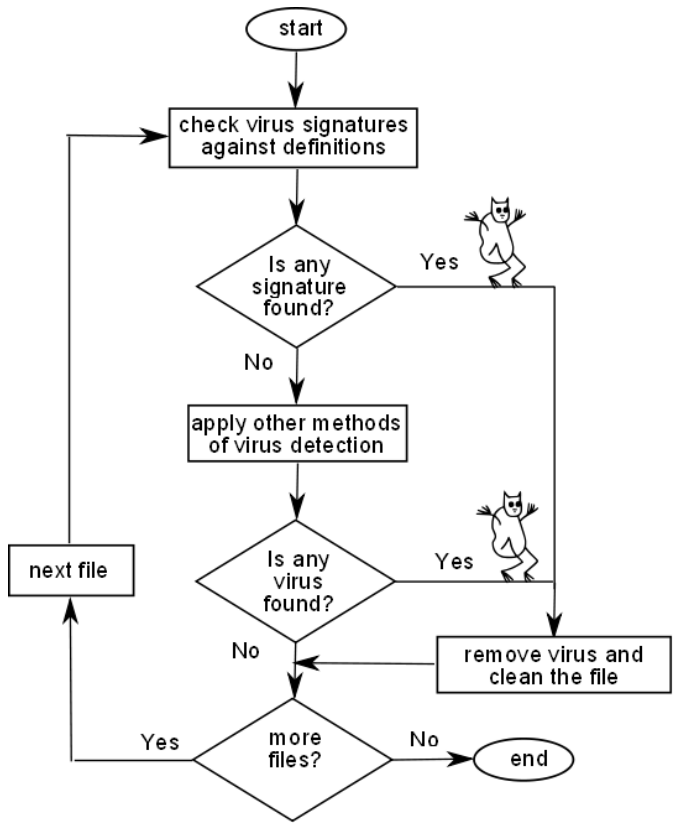

An anti-virus first tries to detect the presence of viruses using different detection methods. Signature detection, being the most popular method, may be applied first. If no signature is found then the anti-virus applies other methods like heuristic scanning. If no suspicion is raised on a file, then the file is deemed to be uninfected. The sequence of virus canning and cleaning is depicted in the flowchart at the left. The anti-virus has to scan all the files one after other and take a series of actions on the infected files.

The first attempt of any anti-virus product is to repair the damaged files or sectors of the disk. However, if the anti-virus does not know the method of repairing the infection then it isolates the infected file to quarantine for a possible repairing in future. If a virus is found to be too dangerous or the file is severely damaged then the anti-virus may decide to delete the infected file. Thus the actions are generally configured in a sequential order, such as, repairing the file (most preferred), quarantining the file (if cannot be repaired) and deleting the file (least preferred).

**Repairing the infected file**
This is obviously the best method. The anti-virus applies various methods to remove the virus code and repair the file to bring it back to its original form. But repairing can be difficult depending on the nature and extent of alteration made by the viruses.

### Restoring original files from a backup

Many operating system files and other critical application files are backed up secretly by the anti-virus. When any of these files are altered by virus infection, the anti-virus deletes the infected files and recovers the original files from the backup. This is a quick and easy method for recovering the original file without knowing the name of the virus or type of infection or method of disinfection. However, this is possible only for limited number of files those are backed up. Not every file can be restored by this method.

### Putting into Quarantine

Sometimes the anti-virus knows that the file is infected but does not know how to disinfect it (nor has a copy in backup). In such cases the anti-virus puts the file into the quarantine with a hope of repairing the file in future. If the vaccine is available to the anti-virus within a reasonable time frame, then file is repaired otherwise deleted after a period.

### Deleting the infected file

This is the worst out of all the options that an anti-virus may resort to. When a file is totally damaged or infected by dangerous viruses then the anti-virus decides to delete it. Deleting a file may result in non-functioning of an application or Operating System.

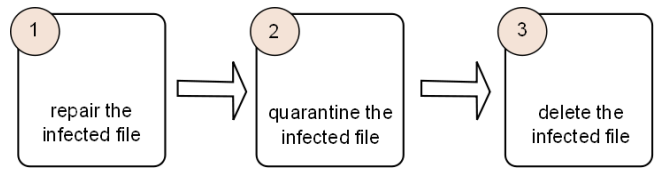

Steps of Virus removal

### TRIZ based analysis

While working on an infected file, the IFR (Ideal Final Result) of an anti-virus is definitely to bring the file back to its original position (TRIZ concept of Ideality). If there is a backup of the infected file then the anti-virus deletes the infected file and restores the original file from the backup (Principle-26: Copying). If a method of repairing the infected file is known to the anti-virus then it applies the method to repair the file (Princple-13: Reversing). If there is no backup and no knowledge of repairing then infection, then the anti-virus sends the infected file to quarantine or deletes the file permanently from the computer (Principle-2: Taking out).

## 3. Methods of Virus removal and file repairing

In case of simple infections, the anti-virus has to first restore the original bytes overwritten by the virus code. Those bytes are usually found somewhere there in the virus code. This is because the virus typically restores those bytes to run the host program after running the virus code. After the cleaner finds those initial bytes, it has to put them back in their original location and remove the appended virus code to truncate the file to its original size.

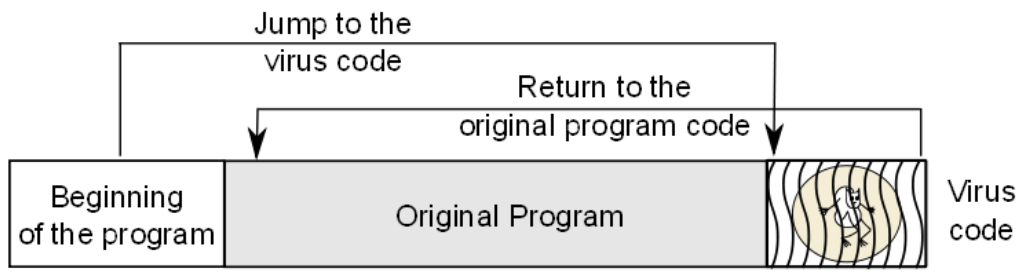

Typical Execution of an Infected Program

- For example, if a system is infected with Jerusalem/PLO virus, the scanner refers to the virus information database to get the pattern of infection and cleaning method. In this case the scanner may get the following information from the virus information database and send the cleaner to take necessary action to clean.

| Virus Name | Virus behavior | Cleaning method |
|---|---|---|
| Jerusalem/ PLO virus | The virus is 1873 bytes in size; it overwrites the first three bytes of the original program with a jump to itself. The original bytes are located at offset 483 in the viral code. | The cleaner has to take three bytes from offset 843 and copy them to the beginning of the file. Then it has to remove 1873 bytes of the file. |

- When a system is infected by a boot sector virus or the system is unable to boot because of damaged boot sector, then the anti-virus has to boot the system from a clean external disk in order to scan the system further. In order to repair boot sector infection usually the infected boot sector is overwritten by an uninfected boot sector.

- While repairing an email the anti-virus may have to first detach the attachments from the email body, then scan the attachments. If no infection is found in the attachments, the email is released as usual. If an infection is found then the attachment is disinfected (or deleted if cannot be disinfected) and then attached back to the email. Then the email is released with the repaired attachments.

- A similar process is followed while disinfecting macro viruses. The infected macros are first detached from the original document for cleaning and again attached to the original document after cleaned.

- If the anti-virus does not have any knowledge of repairing the file then it has to delete the file or move the file to quarantine. It may repair the files in quarantine at a later period when it gains knowledge about repairing the infection.

## 4. Drawbacks of the conventional methods of cleaning

In case of simple infections the original program code remains in one single block. But there are situations where the virus may scramble the original program. These situations are very challenging for the anti-virus program to repair. The following are some difficulties of cleaning virus infections.

> There are some infections where parts of the files are damaged by the virus. These types of infections are caused by "file modifying viruses". In these cases, the chance of recovery is less, but the anti-virus has to apply various methods with hope.

> Removing virus is risky and error prone as the removal tool has to manipulate binary executable files (except in macro viruses). A single byte mistake in a binary code can lead to dangerous results like file corruption, application crash or system halt.

> The virus cleaner must know the characteristics of a virus in order to remove that virus. It cannot remove an unknown virus whose methods of infection are not known.

> Beside the actual virus should be exactly the same as the virus detected by the scanner. If a virus is wrongly detected to be a different virus, then the cleaner will do wrong operations and build a garbage file.

- Some viruses have different family members with different features. For example, the Jerusalem/PLO family now contains more than 100 mutants. In such cases the cleaner must exactly determine the virus type in order to clean it successfully.

- Many polymorphic and metamorphic viruses have variable lengths. In such cases the virus length cannot be available from virus definition database. Hence, the conventional cleaners cannot clean such infections.

- If the files are actually damaged by virus infection then simply removing of virus code cannot restore the original program.

- If the malware is a Trojan horse, then there is nothing to clean or restore. As the malicious codes are built into the program, the only option is to delete the program completely.

- Similarly in case of rootkits no attempt is made to repair the files. The only option is to reinstall the programs or restore from healthy backups.

## 5. Advantages of Heuristic Cleaning

The heuristic cleaning first understands the infection mechanism and uses the same knowledge to clean the file. The most interesting part of a virus is that it has to pass the control to the original program after execution of the viral code (except overwriting viruses, which cannot be cleaned in any way). In order to execute the original program every virus must be able to fix up the first few bytes and repair the original program. This information is extremely valuable to repair the infected file.

**Let the Virus do the Dirty Work**
Frans Veldman, the author of TBAV, describes the method of heuristic cleaning in a beautiful way (http://mirror.sweon.net/madchat/vxdevl/vdat/epheurs1.htm). If the virus knows how to repair the infected file then why not to let the virus do the repairing job. The concept of heuristic cleaning is based on this principle. The heuristic cleaner loads the program file in the virtual machine and starts emulating the program code. The method tracks what the virus is doing. When the virus jumps back to the original program code the cleanser stops the emulation process and restores the original bytes from the repaired program. The repaired beginning of the program is then copied back to the infected program file on the disk to repair the program.

Let's analyze the situation from TRIZ perspective. We find that we have the following resources available with us. (i) the infected file, (ii) the virus impressions, and (iii) the virus infection behavior. What we want is to reconstruct the original files from the infected files. This method of disinfection is just the reverse of the infection method (Principle-13: Reversing).

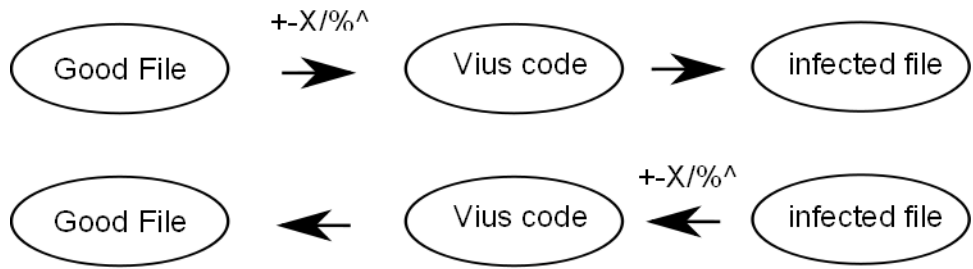

This method is extremely different from the conventional cleaning as it is "removing the unknown from the unknown". This is a miraculous job as it does the cleaning without knowing the name of virus or the nature of infection. However, there is a possibility of many things going wrong. For example;

> If the virus does not intend to execute the original program this heuristic driver may take you to a wrong place. In this case the purpose of restoring the original program fails.

> The virus may not let you go out of the viral code by putting you in an endless loop. In this case the heuristic cleaner has to wait forever.

> If the virus does not return the control to the original program the cleaner will be too ambitious to emulate further. In that case the virus gets control of the emulated environment and finally escapes from it.

> If the virus is badly programmed, it may crash during emulation thereby crashing the emulator too.

## 6. Invention on Restoring Virus Infected Systems

Sometimes the computers are not able to boot because of virus infection. Unless the computers boot the anti-virus cannot be loaded to scan the system. It is necessary to restore a damaged computer and provide up-to-date virus scanning. There are some innovative methods on restoring an infected system.

> US Patent 6347375 (invented by Reinert et al., assignee Ontrack Data International, Feb 2002) provides a remote scanning and repairing mechanism contrary to the conventional method of local scanning and repairing. According to the invention the local computer has to boot from a bootable virus utility operating program and establish communications with a remote computer. Then the local computer downloads a remote data recovery program from the remote computer to the memory of the local computer and executes the data recovery program in the memory of the local computer. Then the local computer downloads a virus scanning and repair utility from the remote computer to the memory of the local computer and executes the virus scanning and repair utility at the local computer.

7) US Patent 7353428 (invented by Cheston, et al., Assignee- Lenovo Singapore Pte. Ltd, Apr. 2008) teaches that the client computer connects to a pre-authorized anti-virus server and informs the server that an anti-virus needs to be immediately downloaded. After that the client computer is disconnected from the network and reconnects a link only with the trusted anti-virus server. This is performed by applying a filter in the network interface card (NIC) driver by the primary OS or the secondary OS or a service processor (SP) or by a virtual machine manager (VMM) depending on which is available at the client computer (Principle-2: Taking out). After a trusted connection is established the anti-virus fix is automatically installed without requiring any end-user action. After the anti-virus fix is installed and the client computer is rebooted, the client computer is allowed to reconnect the full network.

## 7. Exterminating Memory Resident Viruses

A large number of computer viruses are memory resident in nature, which monitor commands of a BIOS or DOS from within the memory. The traditional virus scanners can only detect a virus in the system area but cannot exterminate a virus that is resident in the memory. Therefore, even if a file is healed from virus, it becomes infected again by the resident virus. A manual extermination process is difficult and error prone. There is a need for a simple and risk-free method to exterminate the memory resident viruses.

| 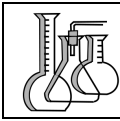 | **INVENTION- Automatic extermination of memory resident viruses** |
|---|---|

Patent 6240530 (invented by Togawa, assignee Fujitsu Limited, May 2001) provides an optimum virus extermination method in four automatic steps. The first step involves virus detection and identification. If a virus is detected then the second step involves memory clearing. The third step involves restarting the system by fetching a clean operating system from outside. After that the virus is exterminated in the fourth step.

This method of exterminating the virus is much easier compared to a manual extermination process. It clears the information in the memory and fetches a virus-free operating system automatically, thus prevents any errors that would have been made by the user during a manual extermination process (Principle-25: Self service).

## 8. Innovative Methods of Quarantining Files

**Contradiction**
There are some infections which cannot be repaired at the time of scanning as because the method of repairing is not known. If the anti-virus deletes those files then the files are gone forever. If the anti-virus does not delete those files then they will infect other files in the computer.

**Solution**
The anti-virus should isolate the infected files to a different directory to repair them at a later stage when the method of repairing will be known (Principle-2: Taking out). The anti-virus can further rename the infected files (such as from .exe to .xyz) in order to prevent them from being used (Principle-35: Change Parameter).

**Further problem**
Although this method prevents spreading of the virus, the virus may still be inadvertently spread if the file is executed or transferred to another processing system. Hence, there is a need for a better method of quarantine to manage the virus-infected files.

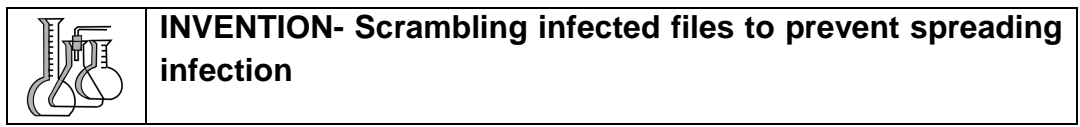
**INVENTION- Scrambling infected files to prevent spreading infection**

Patent 6401210 (Method of managing computer virus infected files, invented by Templeton, assignee Intel Corporation, June 2002) suggests to scramble the virus infected files and safely store them in a special directory like virus bin. The scrambled infected file rules out the possibility of being used. The invention uses an unscrambler logic to unscramble the virus infected file to reproduce the virus infected file in future if necessary.

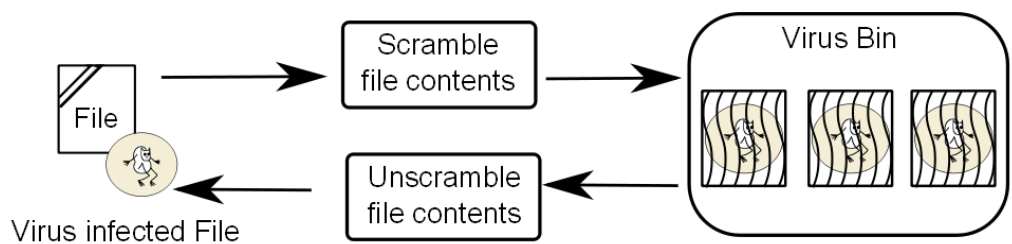

The invention uses scrambling techniques to make the infected files unusable (Principle-36: State Change or Conversion) and isolates them to a virus bin (Principle-2: Taking out).

## 9. Inventions on Repairing the Infected Files

As we have discussed earlier one option of restoring the original file is to copy it back from a backup or install again from its original source. But there are situations where there is no backup of the file and the original source is either not known or not available. In all these cases the anti-virus must apply all its intelligence to rebuild the file. We have already discussed various method of repairing an infected file including the powerful heuristic method which can repair the unknown infection by unknown viruses. But there are still some difficulties.

**Further Problems**
Although a reverse application of the virus code and logic should produce back the original file, there are some problems in this process. One problem is that if the file is actually damaged by the virus then it cannot probably be reconstructed by applying whatever method. Secondly even if the files is reconstructed and restored, we cannot ensure that the reconstructed file is same as the original file.

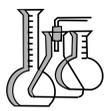 **INVENTION- Method for recovery of a computer program infected by a computer virus (Patent 5408642)**

Patent 5408642 (Invented by Omri Mann, assignee Symantec Corporation, April 95) proposes a recovery technique by repairing and reconstructing the infected file. The invention generates a unique fingerprint for each computer program before any file is infected. Then it stores the unique fingerprint, the data relating to the beginning portion of a program and the length of the program at a separate location prior to the program being infected (Principle-2: Taking out, Principle-26: Copying).

Later when the program is found to be infected by comparing the fingerprints, the program is reconstructed by using the beginning portion of the program. A new fingerprint is generated from the repaired program and is matched with the fingerprint taken before the infection. If the fingerprints match then the infected file is replaced by the repaired program. Thus the files are reconstructed and verified without backing up and restoring huge files.

As we have discussed earlier, every virus do not infect the files in the same way. Hence, treatment of the files also has to be different. For example, the conventional method of deleting infected .exe files is not suitable for treating macro viruses. Macro viruses generally infect document files and the files might contain useful information. Deleting the infected file or macro would result in loss of that information.

**INVENTION- System, apparatus and method for the detection and removal of viruses in macros**

Patent 5951698 (invented by Chen, et al.) proposed a method of decoding, scanning and treating macro viruses. The invention includes a macro locating and decoding module, a macro virus-scanning module, a macro virus treating module, a file correcting module and a virus information module.

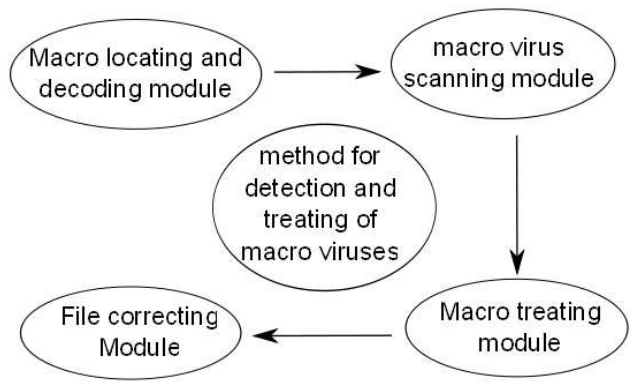

First the macro-locating module determines whether the targeted file includes a macro. If a macro is found, it is decoded and stored in the data buffer (Principle-2: Taking out). Then the scanning module accesses the decoded macro and scans it to determine whether it includes any viruses. First the decoded macro is scanned for known viruses. If a known virus is found then the macro is flagged as infected. If a known virus is not detected the scanning module further determines whether the decoded macro includes an unknown virus.

Then the macro-treating module removes the suspect instructions from the infected macro to produce a cleaned or sanitized macro (Principle-2: Taking out). Finally the file-correcting module replaces the infected macro with the treated macro to produce the clean and virus free file (Principle-5: Merging).

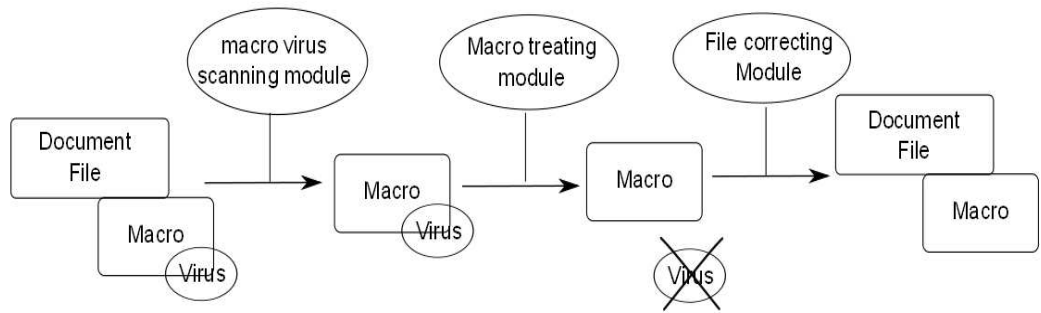

**A TRIZ based analysis:**

Although the scanning for viruses was known to the prior art, this invention makes it specific for the macro viruses, as the method of treating macro viruses is different from treating other viruses. The invention first detaches the macro from the file (Principle-2: Taking out) then detaches the virus from the macro (Principle-2: Taking out) and finally joins the treated macro to the original file (Principle-5: Merging). Thus the invention provides a comprehensive solution using four interlinked modules (Principle-40: Composite).

/١) Patent 7512808 (invented by Liang, Assignee-Trend Micro, Inc. (JP), Mar. 2009) discloses how to remove viruses from computers by using anti-computer virus agent. The invention creates an anti-computer virus agent by parsing and modifying the virus code. This modified virus detects whether a client computer is infected with that virus and in case infected, performs a reverse action for cleaning and repairing of the damages done to the client device (Principle-13: Other way round).

# 10. Inventions on Restoring Infected Files from Backup

This method of virus cleaning maintains a backup of the uninfected files. If the files are found corrupted due to hardware problem or virus infection, then the files are restored from the previously stored backup. However there are some drawbacks in this method.

- 👎 In order to implement backup mechanism for a large number of files the method requires huge storage space and voluminous backup restore operations.

- 👎 Even if the files are restored from the backup, the modifications that have been done since the last backup cannot be recovered.

- 👎 If the backed up data contains a virus, then the system may still remain contaminated and the files may again be corrupted. There is a need to recover files that have been modified since the last backup and ensure that the restored files don't contain any viruses or worms.

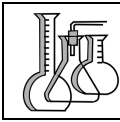 **INVENTION- Safe restoring of un-backed up data from a failing system**

Patent 7222143 (invented by Childs, et al., assignee- Lenovo Pte Ltd., May 2007) discloses a method of restoring a previously un-backed up data during a system restore.

The invention uses a locked partition with an alternate operating system. The locked partition should be accessible only by the alternate operating system and not accessible by the primary operating system.

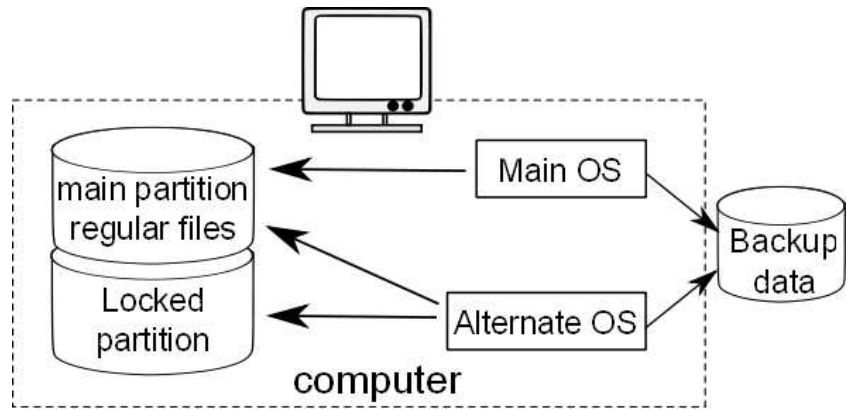

While restoring the failing system, the alternate operating system determines which files have been modified since the most recent backup and runs a virus scan on those modified files and uncorrupts the corrupted files. Thus the alternate operating system recovers all the modified files from the main partition to the locked partition, either uninfected or disinfected by the virus scanner. After the data restored from the last backup, the files which have been modified since last backup are replaced with the files recovered in the locked partition. Thus the system recovers files since the most recent backup while ensuring at least part of the restored files don't contain any viruses. Let's examine the application of TRIZ inventive standards in this case.

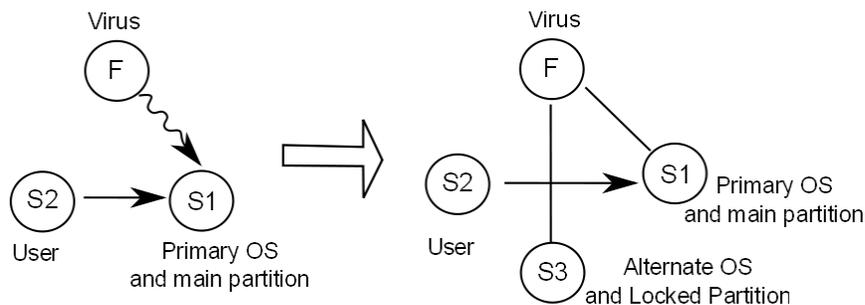

Introducing an alternate OS
and a Locked Partition
Inventive Stadard 1.2.1

According to Inventive Standard 1.2.1, if there is a harmful interaction between two substances and there is no need to maintain a direct contact between the substances, the problem can be solved by introducing a third substance between them. In this case virus has a harmful effect on the main operating system and the main partition of the disk. To invention introduces a third substance, i.e., an alternate OS and a locked partition to provide a solution.

| 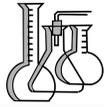 | **INVENTION- Restoration of data corrupted by viruses using pre-infected data** |

Patent **7392542** (invented by Bucher, Assignee- Seagate Technology LLC, Jun. 2008) discloses a method of restoring pre-infected data in a network by using data mirroring. According to the invention a network appliance keeps a mirrored backup of all the computers in the network in an ongoing manner. There will be an anti-virus module in the network appliance which will scan the data received from individual computers. If no virus is detected, then the data is updated in the mirror. If on the other hand, a virus is detected, the mirrored copy is not updated. Instead, the previous copy from the mirror, representing a pre-infection state of the data, is used to restore the infected computer.

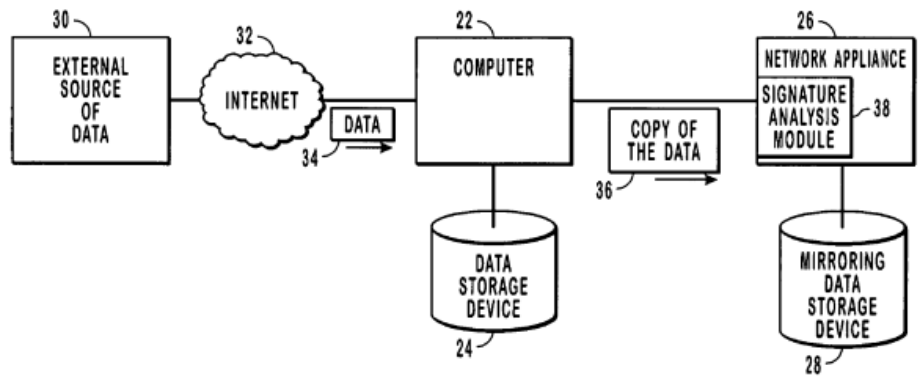

In this manner, a single copy of an anti-virus utility executed at the network appliance can protect a large number of computers in a network, without requiring individual copies of anti-virus software for each machine.

# Reference:

1. Umakant Mishra, "TRIZ Principles for Information Technology", Revised Edition 2010

2. Umakant Mishra, "Using TRIZ for Developing Anti-Virus", Book Under Publication

3. Yuri Salamatov, "TRIZ: The right solution at the right time", 1999

4. Altshuller G., "Creativity as an Exact Science: The Theory of the Solutions of Inventive Problems", Gordon & Breach, 1988.